\documentclass[a4paper]{article}
\usepackage[T1]{fontenc}
\usepackage{graphicx}
\usepackage{amssymb}
\usepackage{amsmath}
\usepackage{amsfonts, amsthm}

\usepackage{url} 

\def\mymedskip{\vskip\medskipamount}
\def\mymedbreak{\par \ifdim\lastskip<\medskipamount
  \removelastskip \penalty-100 \mymedskip \fi}
\def\myaftermedspace{\par \ifdim\lastskip<\medskipamount
  \removelastskip \penalty55\mymedskip\fi}
\newcommand{\eop}{{\unskip\nobreak\hfil\penalty50
          \hskip2em\hbox{}\nobreak\hfil$\Box$
          \parfillskip=0pt \finalhyphendemerits=0 \par}}
{\mymedbreak{\noindent\bf Proof #1.\enspace}}{\eop\myaftermedspace}
{\mymedbreak{\noindent\bf Proof of Theorem~\ref{#1}:\enspace}}{\eop\myaftermedspace}

\newtheorem{teor}{Theorem}[section]
\newtheorem{defi}[teor]{Definition}
\newtheorem{fct}[teor]{Fact}
\newtheorem{prob}{Problem}
\newtheorem{exercise}{Exercise}
\newtheorem{examp}[teor]{Example}
\newtheorem{lem}[teor]{Lemma}
\newtheorem{cor}[teor]{Corollary}
\newtheorem{con}[teor]{Conjecture}
\newtheorem{prop}[teor]{Proposition}
\newtheorem{rem}[teor]{Remark}

\newcommand{\beq}{\begin{equation}}
\newcommand{\eeq}{\end{equation}}
\newcommand{\beql}[1]{\begin{equation} \label{#1}}
\newcommand{\eeql}{\end{equation}}
\newcommand{\beqa}{\begin{eqnarray*}}
\newcommand{\eeqa}{\end{eqnarray*}}
\newcommand{\beqal}[1]{\begin{eqnarray} \label{#1}}
\newcommand{\eeqal}{\end{eqnarray}}
\newcommand{\beqan}{\begin{eqnarray}}
\newcommand{\eeqan}{\end{eqnarray}}
\newcommand{\bpf}{\begin{proof}}

\newcommand{\epf}{\end{proof}}
\newcommand{\ben}{\begin{enumerate}}
\newcommand{\een}{\end{enumerate}}
\newcommand{\bit}{\begin{itemize}}
\newcommand{\eit}{\end{itemize}}

\newcommand{\bab}{\begin{abstract}}
\newcommand{\eab}{\end{abstract}}
\newcommand{\bke}{\begin{keywords}}
\newcommand{\eke}{\end{keywords}}

\newcommand{\btm}[1]{\begin{teor} \label{#1}}
\newcommand{\etm}{\end{teor}}
\newcommand{\btmn}[2]{\begin{teor}[#1] \label{#2}}
\newcommand{\etmn}{\end{teor}}
\newcommand{\ble}[1]{\begin{lem} \label{#1}}
\newcommand{\ele}{\end{lem}}
\newcommand{\bLe}[1]{\begin{Lemma} \label{#1}}
\newcommand{\eLe}{\end{Lemma}}
\newcommand{\bpn}[1]{\begin{prop} \label{#1}}
\newcommand{\epn}{\end{prop}}
\newcommand{\bex}[1]{\begin{examp} \label{#1}}
\newcommand{\eex}{\hfill$\Box$\end{examp}}
\newcommand{\bde}[1]{\begin{defi} \label{#1}}
\newcommand{\ede}{\end{defi}}
\newcommand{\bco}[1]{\begin{cor} \label{#1}}
\newcommand{\eco}{\end{cor}}
\newcommand{\bcorn}[2]{\begin{cor}[#1] \label{#1}}
\newcommand{\ecorn}{\end{cor}}
\newcommand{\bcon}[1]{\begin{con} \label{#1}}
\newcommand{\econ}{\end{con}}
\newcommand{\bfct}[1]{\begin{fct} \label{#1}}
\newcommand{\efct}{\end{fct}}
\newcommand{\bpr}[1]{\begin{prob} \label{#1}}
\newcommand{\epr}{\end{prob}}
\newcommand{\bexer}[1]{\begin{exercise} \label{#1}}
\newcommand{\eexer}{\end{exercise}}
\newcommand{\bre}[1]{\begin{rem} \label{#1}}
\newcommand{\ere}{\end{rem}}
\newcount\tpcnt
\newenvironment{tproblem}{%
  \global\advance\tpcnt1%
  \goodbreak\medskip\par\noindent\textbf{Problem~\the\tpcnt.}~}%
{%
  \goodbreak
}

\newenvironment{Solution}[1][]{%
  \goodbreak\smallskip\par\noindent\textbf{Solution{\if#1\empty\else~#1\fi}.}~}%
{%
  \goodbreak
}

\newcommand{\cP}{{\cal P}}

\newcommand{\x}{{\bf x}}
\newcommand{\y}{{\bf y}}

\newcommand{\bfc}{\mathbf{c}}

\newcommand{\PG}{{\rm PG}}

\newcommand{\gb}{\beta}

\newcommand{\gd}{\delta}

\newcommand{\gre}{\epsilon}

\newcommand{\gl}{\lambda}

\newcommand{\Tm}[1]{Theorem~\protect\ref{#1}}

\newcommand{\Ex}[1]{Example~\protect\ref{#1}}

\newcommand{\Sec}[1]{Section~\protect\ref{#1}}

\newcommand{\bbF}{\mathbb{F}}

\newcommand{\bbZ}{\mathbb{Z}}

\newenvironment{hint}{\noindent {\bf Hint:} \enspace}{\eop\myaftermedspace}

\newenvironment{multisolution}[1]{\noindent {\bf Solution #1:} \enspace}{\eop\myaftermedspace}

\newcommand{\bqu}{\begin{question}}
\newcommand{\equ}{\end{question}}
\newcommand{\bs}{\begin{solution}}
\newcommand{\es}{\end{solution}}
\newcommand{\bh}{\begin{hint}}
\newcommand{\eh}{\end{hint}}
\newcommand{\bms}[1]{\begin{multisolution}{#1}}
\newcommand{\ems}{\end{multisolution}}

\newcommand{\btp}{\begin{tproblem}}
\newcommand{\etp}{\end{tproblem}}
\newcommand{\bts}{\begin{Solution}}
\newcommand{\ets}{\end{Solution}}

\newcommand{\bfnull}{{\bf 0}}

\newcounter{penumi}
\newenvironment{pit}{%
\begin{list}{(\roman{penumi})}{\usecounter{penumi}\setlength{\labelwidth}{1cm}\setlength{\itemindent}{0pt}\setlength{\topsep}{0pt}\setlength{\parsep}{0pt}\setlength{\partopsep}{0pt}\setlength{\itemsep}{0pt}}
} 
{\end{list}}
\newcommand{\bpit}{\begin{pit}}
\newcommand{\epit}{\end{pit}}

\newcommand{\la}{\langle}
\newcommand{\ra}{\rangle}

\newcommand{\choice}[5]{
\left\{ \begin{array}{ll} #1, & \mbox{#2};\\
                                   #3, & \mbox{#4}#5
\end{array}
\right. 
}
\newcommand{\twochoice}[5]{\choice{#1}{#2}{#3}{#4}{#5}}

\newcommand{\bfaa}{\boldsymbol{a}}

\newcommand{\bfh}{{\bf h}}

\newcommand{\GG}{\boldsymbol{G}}
\newcommand{\bbb}{\boldsymbol{b }}
\newcommand{\cc}{\boldsymbol{c}}
\newcommand{\e}{\boldsymbol{e}}
\newcommand{\g}{\boldsymbol{g}}
\newcommand{\h}{\boldsymbol{h}}
\newcommand{\ii}{{\boldsymbol{i}}}

\newcommand{\s}{\boldsymbol{s}}
\newcommand{\uu}{\boldsymbol{u}}

\newcommand{\aaa}{\boldsymbol{a}}
\renewcommand{\x}{\boldsymbol{x}}
\renewcommand{\y}{\boldsymbol{y}}

\newcommand{\n}{\boldsymbol{0}}

\newcommand{\Fqkstar}{\bbF_q^k\setminus\{\bfnull\}}
\newcommand{\Fqkonestar}{\bbF_q^{k-1}\setminus\{\bfnull\}}
\begin{document}
\title{PIR Codes, Unequal-Data-Demand Codes, and the Griesmer Bound}
%
%
\author{Henk D.L.~ Hollmann\thanks{Hollmann $^{[0000-0003-4005-2369]}$ is with the Institute of Computer Science, University of Tartu, Tartu 50409, Estonia. Email: henk.hollmann@ut.ee}
\and
Martin Pu\v{s}kin\footnotemark\addtocounter{footnote}{-1}
\and
Ago-Erik Riet%
\thanks{Pu\v{s}kin and Riet $^{[0000-0002-8310-6809]}$ are with the Institute of Mathematics and Statistics, University of Tartu, Tartu 50409, Estonia. 
Email: \{martin.puskin,ago-erik.riet\}@ut.ee}
}
%
%
%
\maketitle              
\begin{abstract}
Unequal Error-Protecting (UEP) codes are error-correcting (EC) codes designed to protect some parts of the encoded data better than other parts. 
Here, we introduce a similar generalization of PIR codes that we call Unequal-Data-Demand (UDD) PIR codes. These codes are PIR-type codes designed  for the scenario where some parts of the encoded data are in higher demand than other parts. We generalize various results for PIR codes to UDD codes. Our main contribution is a new approach to the Griesmer bound for linear EC codes involving an Integer Linear Programming (ILP) problem that generalizes to linear UEP codes and linear UDD PIR codes.

\end{abstract}
\section{\label{LSint}Introduction}
%
%
A  $t$-PIR code stores a data record in encoded form on a collection of servers in such a way that the data symbol in any position in the record
can be recovered from the encoded data symbols stored by any of $t$ disjoint groups
of servers; such a group of servers is called a {\em recovery set\/} for that position. We refer to~\Sec{LSUDD} for a more formal definition of PIR codes.

A Private Information Retrieval (PIR) scheme stores a database in encoded form on a multi-server distributed data storage system in such a way that a user can extract a bit of information from the database without leaking information about which particular
bit the user was interested in. 
Originally, PIR codes were employed to reduce the amount of storage needed to implement such a system. Here, linear $t$-PIR codes can be used to implement a classical (linear) $t$-server PIR scheme \cite{CKG+95} with less storage overhead than the original scheme, by using the PIR code to emulate the~$t$ servers \cite{FVY15}; see also~\cite{var-you} for a nice explanation of how PIR codes can achieve this.

Unequal-error-protecting or UEP codes are error-correcting codes that protect some parts of the encoded data better than other parts. A simple example of an UEP code can be obtained by the concatenation of two codes with different error-correcting capabilities. Interestingly, there exist UEP codes that are more efficient than any code obtained by concatenation of smaller codes, see, e.g., \cite{DR78},~\cite[Chapter 1]{Gil88}.  
In analogy with UEP codes, we define {\em Unequal Data Demand\/} or UDD codes as PIR codes designed for cases where some parts of the data are more in demand, more popular, than other parts.  Again, the basic question is whether we can do better than just using a concatenation of two PIR codes with different values for $t$? It turns out that this question again has an affirmative answer and in this paper we will give several examples to show this.

The Griesmer bound \cite[Chapter 17, Theorem 24]{MS77} is a famous and fundamental bound on the length of a linear error-correcting (EC) code with a given minimum distance (see \Sec{LSECC} for definitions). This bound has been generalized to UEP codes in \cite[Chapter 1]{Gil88}, and to PIR codes in~\cite{KY21}. In this paper, we generalize the Griesmer bound to the case of UDD PIR codes. First we show that the Griesmer bound for UDD PIR codes can be obtained as a consequence of the corresponding bound for UEP codes; to this end, we first generalize a well-known bound for the minimum distance for PIR codes \cite{HL22}, \cite{LR18}, \cite{LS14}, \cite{Ska18}, \cite{ZS15} to the case of UDD codes. We also provide an alternative, direct proof, using an Integer Linear Program (ILP) formulation. Interestingly, we show that the ILP can be used to provide a uniform proof for {\em all\/} the Griesmer bounds mentioned above.

The contents of this paper are the following. In Sections~\ref{LSECC}, \ref{LSPIR}, and \ref{LSUEP}, we briefly review error-correcting codes, PIR codes, and UEP codes, respectively. The new notion of UDD PIR codes is introduced and discussed in~\Sec{LSUDD}. In \Sec{LSgries}, we first derive the new distance bound for UDD PIR codes, which we then use to give a first proof of the Griesmer bound for UDD PIR codes. In \Sec{LSILP} we derive an ILP bound for UDD PIR codes and use it to generalize a bound for PIR codes from~\cite{KY21}. We determine a lower bound to the optimum of the ILP in~\Sec{LSsol}, which then provides a second proof of the Griesmer bound for this type of codes. In \Sec{LSuni} we show that our ILP bound provides a uniform proof of the Griesmer bound for all codes mentioned earlier. A few open problems are discussed in~\Sec{LSopen}. Finally, in~\Sec{LScon} we present some conclusions. This work is based on~\cite{Pus22}, where many other results for PIR codes are generalized to UDD codes.

Throughout this paper, $\bbF_q$ denotes the finite field of order $q$, where $q$ is a power of a prime~$p$, and we write $\bbF_q^*=\bbF_q\setminus \{0\}$. We let  $\bbF_q^n$ denote the $n$-dimensional vector space over~$\bbF_q$. For $\bfaa, \bbb\in \bbF_q^n$, we let $\la\bfaa, \bbb\ra:=a_1b_1+\cdots a_nb_n$ denote
the {\em inner product\/} of $\bfaa$ and $\bbb$, see, e.g., \cite{Lin99}. For $\h\in \bbF_q^n$, we let $\h^\perp:=\{\x\in \bbF_q^k\mid \la \x,\h\ra=0\}$ denote the hyperplane with normal vector~$\h$. We use the set $[n]:=\{1, 2, \ldots, n\}$ to index the symbols in vectors (or codewords)  $\cc\in \bbF_q^n$, that is, the symbol in~$\cc$ with index $i\in[n]$ is $c_i$. We use $\bbZ_+$ to denote the set of nonnegative integers.

\section{\label{LSECC}
Error-correcting codes}
The {\em Hamming weight\/} $w(\cc)$ of a vector $\cc\in\bbF_q^n$ is the number of positions $i$ in~$[n]$ for which $c_i\neq 0$, and the {\em Hamming distance\/} $d(\x,\y)$ between $\x, \y\in\bbF_q^n$ is defined as $d(\x,\y)=w(\x-\y)$, that is, the number of positions in which $\x$ and $\y$ differ. 
Note that the Hamming distance is a {\em metric\/}, see, e.g.,  \cite{MS77}.
%
The minimum (Hamming) distance $d(C)$ of a set $C\subseteq \bbF_q^n$ 
is the minimum Hamming distance $d(\x,\y)$ between two {\em distinct\/} vectors $\x, \y\in C$. 

A $(n,M,d)_q$ code $C$ is a set of $M$ vectors from $\bbF_q^n$ with minimum distance $d(C)= d$. The elements of~$C$ are referred to as {\em codewords\/}. We say that $C$ is {\em linear\/} if $C$ 
is a 
linear subspace of~$\bbF_q^n$; if $\dim(C)=k$ then we say that $C$ is a $[n,k]_q$ code, or a $[n,k,d]_q$ code if $d(C)=d$.
A {\em generator matrix\/}~$\GG$ for a linear code~$C$ is a $k\times n$ matrix with entries from~$\bbF_q$ with the property that the rows of~$\GG$ consist of codewords that together form a {\em basis\/} for (the $\bbF_q$-linear subspace) $C$. An {\em encoder\/} for~$C$ is a one-to-one map $\gre: \bbF_q^k\rightarrow C$. Given a generator matrix~$\GG$ for~$C$, the map $\gre: \bbF_q^k\rightarrow \bbF_q^n$ defined by $\gre(\aaa)= \aaa^\top \GG$ is a {\em linear\/} encoder for~$C$, referred to as {\em the encoder $G$\/} for~$C$;
note that any linear encoder is of this form. 
\section{\label{LSPIR}PIR codes}
We first provide a formal definition of a PIR code. 
\bde{LDpir}Given a (one-to-one) encoder map $\gre: \bbF_q^k \longrightarrow \bbF_q^n$, a set of positions $I=\{i_1, \ldots, i_s\}\subseteq [n]$ is called a {\em recovery set for the $j$-th data symbol\/} if the restriction $\bfc_I=(c_{i_1}, \ldots, c_{i_s})$ of a codeword $\bfc=\gre(\bfaa)$ uniquely determines the $j$-th data symbol $a_j$. The encoder map $\gre$ is a $t$-PIR code if  there exists for every $j=1, \ldots, k$ a collection of $t$ disjoint recovery sets for the $j$-th data symbol.

We say that a $k\times n$ matrix $\GG$ with entries from~$\bbF_q$ is a (linear) $t$-PIR code if the corresponding encoder $\gre: \bfaa^\top \rightarrow \bfaa^\top \GG$ is $t$-PIR. In that case we say that~$\GG$ {\em generates\/} a $t$-PIR code, or that $\GG$ is {\em $t$-PIR\/}.
\ede
Here it is important to realize that the $t$-PIR property is a property of the {\em encoder\/} of the code. 
Note that if the span of the columns from a $k\times n$ matrix~$\GG$ with indices in a set of positions~$I$ contains the $j$-th unit vector $\e_j$ ($j\in [k]$), then~$I$ is a recovery set for the $j$-th data symbol in the PIR code generated by~$\GG$. In \cite[Theorem 1]{LS14} it was shown that every recovery set~$I$ arises in this way. 
\bex{LEpirex}\rm Let $q=2$, and let $C$ be the binary linear code with (linear) encoder $\gre: \bfaa^\top \rightarrow \bfaa^\top \GG$, where 
\beql{LEGmat}G=
\left(
\begin{array}{cccc}
1&0&1&1\\
0&1&1&0
\end{array}
\right).
\eeql
Then the first data symbol has recovery sets $\{1\}, \{2,3\}, \{4\}$ and the second data symbol has recovery sets $\{2\}, \{1,3\}$, so this encoder and this matrix $\GG$ are both 2-PIR. Indeed, note that if $\cc^\top=\bfaa^\top \GG=(a_1, a_2,a_1+a_2,a_1)$, then for example $a_1$ can be recovered as $c_2+c_3$, in accordance with the fact that~$\e_1$ is the sum of the second and third column of~$\GG$.
Since every recovery set for the second data symbol except~$\{2\}$ has size at least 2, that data symbol does not have 3 disjoint recovery sets and so $\GG$ is not a 3-PIR code.
\eex
\section{\label{LSUEP}Unequal-Error-Protection codes}
An error-correcting code is designed to protect data against the occasional occurrence of errors: by sending the data in encoded form, the original data can still be recovered from the received codeword as long as not too many errors have occurred. Unequal-Error-Protecting codes play a similar role, but are designed to protect certain data symbols better than others, see, e.g.,  \cite{DR78}, \cite[Chapter 1]{Gil88}.

For an encoder map $\gre: \bbF_q^k\rightarrow \bbF_q^n$, define the {\em separation vector\/} $\s(\gre)\in \bbZ_+^k$ by defining for every $j\in [k]$
\beql{LEsep} s_j(\gre)=\min \{d(\gre(\aaa), \gre(\aaa')) \mid \bfaa, \bfaa'\in\bbF_q^k, a_j\neq a'_j\}.
\eeql
So $s_i(\gre)$ is just the minimal distance between two codewords in distinct sub-codes $C_{i,\gb}=\{\cc=\gre(\aaa) \mid a_i=\gb\}$ ($\gb\in \bbF_q$) of the code $C=\gre(\bbF_q^k)$.
It is not difficult to see that by decoding to the nearest codeword, we can decode the $i$-th data symbol correctly if at most $\lfloor (s_i(\gre)-1)/2\rfloor$ errors have occurred. For more details, see, e.g., \cite[Chapter 1, Section II]{Gil88}.
We note that
\[ d(C)=\min _{j\in [k]} s_i(\gre).\]
We will write $\s(\GG)$ to denote the separation vector of a linear code encoded with generator matrix $\GG$. 

A ``trivial'' construction of an UEP code is to use an $(n,q^{k_1},d_1)_q$  code $C_1$ 
to protect part of the data, and a $(n_2, q^{k_2},d_2)_q$ code $C_2$ 
to protect another part of the data. 
Then the {\em concatenation\/} of $C_1$ and $C_2$, the code with codewords $(\cc_1,\cc_2)$ with $\cc_i\in C_i$ ($i=1,2$) 
has a separation vector $\s(\gre)$ for which 
\[s_i(\gre)\geq\twochoice{d_1}{if $i$ is among the first $n_1$ positions}{d_2}{if $i$ is among the last $n_2$ positions}{.} 
\]
However, often one can do better than this trivial construction.
\bex{LEuep1}\rm
Suppose we want to protect two data bits against errors or erasures, and we want to realize a separation vector $\s(\gre)=(3,2)$. For the trivial construction, we would need two repetition codes, one of length~3 and one of length~2, with an encoder $\gre(ab)=aaabb$ ($a,b\in \bbF_2$). So the minimum length of a ``trivial'' construction would be 5. Now consider the linear PIR code of length~4 generated 
by the matrix~$\GG$ as in (\ref{LEGmat}).
Here again $\s(\GG)=(3,2)$, but now with a code of length only~4. 
\eex
It turns out that a given $k$-dimensional linear code $C$ has an {\em optimal\/} generator matrix $\GG^*$, in the sense that the separation vector $\s(\GG^*)$ of the encoder determined by~$\GG^*$ 
is componentwise larger than or equal to the separation vector $\s(\GG)$ of any other generator matrix~$\GG$ of the code, that is, $s_j(\GG^*)\geq s_j(\GG)$ for every $j\in [k]$, see \cite{DR78}, \cite[Chapter 1, Section II]{Gil88}.
This allows us to speak of~$\s(\GG^*)$ as {\em the separation vector of the code~$C$\/}. Such a matrix~$\GG^*$ can be obtained by a greedy construction, where the first row of~$\GG^*$ is a vector from~$C$ with minimum weight and each further row is a vector from~$C$ of minimum weight outside the span of the rows already chosen. For further details, see~\cite{DR78}.
\section{\label{LSUDD}Unequal-Data-Demand codes}
$t$-PIR codes are designed so that up to $t$ users can obtain each a particular data symbol from data that is stored in encoded form on a number of servers, where every server can be read off at most once. 
Unequal-Data-Demand (UDD) codes enable a similar scenario, but now for the situation where some parts of the data are in higher demand than other parts. We first present a formal definition.
\bde{LDTPIR}Let $T=(t_1, \ldots, t_k)$ where $t_1, \ldots, t_k$ are integers with $t_1\geq t_2\geq\cdots \geq t_k\geq 0$. An UDD $T$-PIR code of length $n$ is an encoder $\gre: \bbF_q^k\longrightarrow \bbF_q^n$ where the $j$-data symbol has at least $t_j$ mutually disjoint recovery sets ($j=1, \ldots, k$). 

We say that a $k\times n$ matrix $\GG$ with entries from~$\bbF_q$ is a (linear) $T$-PIR code if the corresponding encoder $\gre: \bfaa^\top \rightarrow \bfaa^\top \GG$ is $T$-PIR. In that case we say that $\GG$ {\em generates\/} a $T$-PIR code.
\ede

As for UEP-codes, we can use the ``trivial'' construction by concatenation to obtain examples of UDD PIR-codes, but often we can do better.

\bex{LExmix1}\rm
The properties of the matrix $\GG$ in~(\ref{LEGmat}) as stated in~\Ex{LEpirex} show that $\GG$ generates a $(3,2)$-PIR code of length~4. To achieve this by concatenation would require a length-3 repetition code and a length-2 repetition code, for a total length equal to 5.
%
\eex

\section{\label{LSgries}The Griesmer bound for UEP PIR codes}
%
It is well-known that for a $t$-PIR code, the minimum distance~$d$ of the associated code satisfies $d\geq t$, see, e.g., \cite{HL22}, \cite{LR18},  \cite{LS14}, \cite{Ska18}, \cite{ZS15}.  
We will need the following generalization.
\btm{LTdistb}Let $C$ be an $(n,q^k,d)_q$-code with encoder $\gre: \bbF_q^k\rightarrow \bbF_q^n$, and let $\gre$ have separation vector~$\s(\gre)$. If $\gre$ is an UDD $T$-PIR code, where $T=(t_1, \ldots, t_{k})$ and $t_1\geq t_2\geq \cdots \geq t_{k}$, then $s_j(\gre)\geq t_j$ for all $j\in [k]$.
\etm
\bpf
Let $j\in [k]$. Since $\gre$ is $T$-PIR, there are $t_j$ mutually disjoint recovery sets $I_1, \ldots, I_{t_j}$ for the $j$-th data symbol. Now let $\bfaa, \bfaa'\in \bbF_q^n$ with $a_j\neq a'_j$, and let $\bfc=\gre(\bfaa)$ and $\bfc'=\gre(\bfaa')$ be the corresponding codewords. For every $i\in [t_j]$, since $I_i$ determines the $j$-th data symbol and since $a_j\neq a'_j$, we must have $\bfc_{I_i}\neq \bfc_{I_i}'$, that is, $\bfc$ and $\bfc'$  differ in a position in~$I_i$. We conclude that $d(\bfc,\bfc')\geq t_j$. Now the claim follows from the definition of $s_j(\gre)$ in~(\ref{LEsep}).
\epf
We will use this result to prove the following generalization of the Griesmer bound. 
%
\btmn{Griesmer for UDD PIR codes}{LTGries}Suppose that the $k\times n$ matrix $\GG$ over~$\bbF_q$ generates a linear UDD $T$-PIR code, where $T=(t_1, \ldots, t_k)$ with $t_1\geq t_2\geq\cdots \geq t_k\geq 0$. Then 
\beql{LEGries}n\geq \sum_{j=1}^k \left\lceil \frac{t_j}{q^{j-1}}\right\rceil.
\eeql
\etmn
\bpf
Suppose that $\s(\GG)=(s_1, \ldots, s_k)$ is the separation vector of the UEP code generated by~$\GG$. Then by the Griesmer bound for linear UEP codes \cite[Chapter I, Part III, Corollary 14]{Gil88} we have that $n\geq  \sum_{j=1}^k \lceil s_j/q^{j-1}\rceil$. By \Tm{LTdistb}, we have $s_j\geq t_j$, hence (\ref{LEGries}) follows immediately.
\epf
%
It would be nice to have an argument that would prove all these Griesmer-type bounds {\em simultaneously\/}, in a {\em uniform\/} way. In the next sections we will provide such an approach.
\section{\label{LSILP}An 
ILP problem related to PIR codes}
%
%
Fix a prime power~$q$. 
%
There is a one-to-one correspondence between the collection of~hyperplanes in~$\bbF_q^k$ and the collection~$\cP_k$ of vectors $\h\in \bbF_q^k\setminus \{\n\}$ of the form $\h=(0, \ldots, 0, 1, \ldots)$, so with the first nonzero entry equal to 1, where a vector $\h\in \cP_k$ corresponds to the hyperplane $\h^\perp:=\{\bfaa\in \bbF_q^k\mid \la \h,\bfaa\ra=0\}$. For later use, note that $|\cP_k|=(q^k-1)/(q-1)$. (Note also that the vectors in~$\cP_k$ are in one-to-one correspondence with the points in the $(k-1)$-dimensional projective geometry $\PG(k-1,q)$, see, e.g., \cite[Appendix B]{MS77}.) For $\bfh\in \cP_k$, define
\[\nu(\h)=\min \{j\in \{1,\ldots, k\} \mid h_j\neq 0\};\]
as a consequence, $h_{\nu(\h)}=1$. We now have the following. 
(Here and below, for less cumbersome notation, we will write 
$\sum_{{\rm Condition}(\ii)}n_{\ii}$
to denote the sum of all numbers $n_\ii$ with $\ii\neq {\bf 0}$ for which $\ii$ satisfies ${\rm Condition}(\ii)$.)
\btm{LTILPcond} {\rm (Cf.\ \cite[Lemma 6]{KY21})}
Let $\GG$ be a $k\times n$ matrix over~$\bbF_q$ that generates an UDD $T$-PIR code, where $T=(t_1, \ldots, t_{k})$ and $t_1\geq t_2\geq \cdots \geq t_{k}$. Suppose that~$\GG$ has $n_\ii$ columns equal to
$\ii$, 
for $\ii\in \bbF_q^k$.
Then for all $\bfh\in \cP_k$, we have
\beql{LEGineq}\sum_{\la\ii,\h\ra\neq 0} n_\ii\geq t_{\nu(\h)} \eeql
\etm
\bpf
To see this, note that if the $j$-th unit vector $\e_j$ is not contained in the hyperplane $\bfh^\perp$ ($j\in [k], \bfh\in \cP_k$), that is, when $h_j\neq 0$, then every set of columns of~$\GG$ whose span contains~$\e_j$ must contain a column outside~$\bfh^\perp$; so by our assumptions on~$\GG$, there are at least $t_j$ columns of~$\GG$ outside~$\bfh^\perp$. Taking $j=\nu(\bfh)$ gives~(\ref{LEGineq}).
\epf
%
So for $T=(t_1, \ldots, t_k)\in \bbZ^k$ with $t_1\geq \cdots \geq t_k\geq 0$, define $\mu(T)$ to
be the solution of the following Integer Linear Programming (ILP) problem:
\beql{LEILP}
{\small
{\rm ILP}(T):
\left\{
\begin{array}{ll}
n_\ii\in \bbZ, n_\ii\geq 0& (\ii\in\Fqkstar)\\
\displaystyle \sum_{\{\ii \mid \la\ii,\bfh\ra\neq 0\}}^{\mbox{}}n_\ii\geq t_{\nu(\h)} &(\h\in \cP_k)\\
\mbox{minimize } n=\displaystyle\sum_{\ii\in\Fqkstar}n_\ii & 
\end{array}
\right.
}
\eeql
Then, according to \Tm{LTILPcond}, if the $k\times n$ matrix~$\GG$ generates a $(t_1, \ldots, t_k)$-PIR code with $t_1\geq \cdots \geq t_k$, then 
$n\geq n-n_{\bf 0}\geq \mu(T)$, where for an optimal solution, we should of course take $n_{\bf 0}=0$.
%
\bex{LexILP}\rm Let $q=2$ and $k=2$, and let $T=(t_1,t_2)\in \bbZ^3$ with $t_1\geq t_2\geq0$. Associating the numbers $1,2,3$ with the vectors $(1,0)$, $(0,1)$, and $(1,1)$, the ${\rm ILP}(T)$ is the problem to minimize $n=n_1+n_2+n_3$, where $n_i \geq 0$ is integer ($i=1, 2,3$), under the conditions 
\beqal{rCl}
n_1+n_3&\geq& t_1\\
n_2+n_3&\geq& t_2\\
n_1+n_2&\geq& t_1,
\eeqal
where the inequalities correspond to the hyperplanes $(1,0)^\perp$, $(0,1)^\perp$, and $(1,1)^\perp$, respectively.
It is not difficult to see that the minimum value for $n$ under these conditions equals $t_1+\lceil t_2/2\rceil$.
%
\eex
It is easy to give a lower bound for~$\mu(T)$.
\bpn{LPLBILP}We have that $\mu(T)\geq \sum_{j=1}^{k}t_j/q^{j-1}$.
\epn
\bpf
If $\ii\neq \n$, then $|\ii^\perp| =q^{k-1}$, hence $|\bbF_q^k\setminus \ii^\perp|=q^k-q^{k-1}=q^{k-1}(q-1)$. So there are $q^{k-1}$ vectors $\h\in \cP_k$ such that $\la\ii,\h\ra\neq 0$.  As a consequence, using~(\ref{LEILP}) we have
\[ \sum_{j=1}^k t_jq^{k-j}=\sum_{\h\in \cP_k}t_{\nu(\h)}
\leq  \sum_{\h\in \cP_k}\sum_{\la\ii,\h\ra\neq 0} n_\ii
= \sum_{\ii\neq \n}\sum_{\h\in \cP_k, \la\ii,\h\ra\neq 0}n_\ii
=\sum_{\ii\in\Fqkstar}q^{k-1}n_\ii,
\]
so $n\geq \sum_{\ii\in\Fqkstar}n_\ii\geq \sum_{j=1}^{k}t_j/q^{j-1}$.
\epf
In the next section, we will provide a better bound for $\mu(T)$.

\section{\label{LSsol}A sharper lower bound for the ILP problem}

Our aim is to prove the following theorem.
\btm{LTILP}Let $\mu(T)$ be the optimal solution of the ILP problem (\ref{LEILP}), where $\GG$ and $T$ are as in~\Tm{LTILPcond}. Then 
\beql{LEG} \mu(T)\geq \sum_{j=1}^{k}\left\lceil \frac{t_j}{q^{j-1}}\right\rceil.\eeql
\etm
\bpf
To prove this, we will use induction on the dimension~$k$.
First note that the theorem obviously holds for $k=1$. 
Assume that the theorem holds for dimension~$k-1$, and suppose that the $n_\ii$ with $\ii\in\Fqkstar$ satisfy the ILP constraints.

First consider the $q^{k-1}$ inequalities in~(\ref{LEILP}) that involve $t_1$. Let $\gd\in \bbZ_+$ and $\bbb\in \cP_k$ with $\nu(\bbb)=1$ be such that for all $\uu\in\cP_k$ with $\nu(\uu)=1$, we have
\beql{LEG0}\sum_{\la\ii,\uu\ra\neq 0} n_\ii\geq \sum_{\la\ii,\bbb\ra\neq 0} n_\ii= t_{1}+\gd.
\eeql
%
Then for every $\uu\in \cP_k$ with $\nu(\uu)=1$, we have that 
$\sum_{\la\ii,\uu\ra\neq 0}n_\ii\geq \sum_{\la\ii,\bbb\ra\neq 0}n_\ii$, 
and hence
\beql{LErestineq}\sum_{\substack{\la\ii,\uu\ra\neq 0\\\la\ii,\bbb\ra= 0}}n_\ii \geq \sum_{{\substack{\la\ii,\bbb\ra\neq 0\\\la\ii,\uu\ra= 0}}}n_\ii.\eeql
Fix $\h'\in \cP_{k-1}$, and set $\h=(0,\h')$. Note that $\h\in \cP_k$ and $\nu(\h)=1+\nu(\h')>1$. For every $\gl\in \bbF_q$, define $\uu_\gl:=\bbb+\gl \h$; note that $\uu_\gl\in\cP_k$ and $\nu(\uu_\gl)=1$, so that (\ref{LErestineq}) holds for $\uu=\uu_\gl$. Note also that $\la \ii,\uu_\gl\ra=\la \ii,\bbb\ra+\gl\la \ii,\h\ra$. Now
\beql{LEtnuheq}
t_{\nu(\h)}\leq \sum_{\la \ii,\h\ra\neq 0}n_\ii=\sum_{\gl\in \bbF_q}\sum_{\substack{\la \ii,\h\ra\neq 0\\\la \ii,\bbb\ra=-\gl\la \ii,\h\ra}}n_\ii=\sum_{\substack{\la \ii,\h\ra\neq 0\\\la \ii,\bbb\ra=0}}n_\ii+\sum_{\gl\in \bbF_q^*}\sum_{\substack{\la \ii,\h\ra\neq 0\\\la \ii,\uu_\gl\ra=0}}n_\ii.
\eeql
Now note that if $\gl\neq 0$ and $\la\ii,\uu_\gl\ra=0$, then $\la \ii,\h\ra=0$ holds if and only if $\la \ii,\bbb\ra= 0$; note also that if $\gl\neq 0$ and  $\la \ii,\bbb\ra=0$, then $\la\ii,\uu_\gl\ra=0$ if and only if $\la \ii,\h\ra=0$. So noting that $\nu(\uu_\gl)=1$ and using (\ref{LErestineq}), we find that
\beql{LEtnuhint}
\sum_{\substack{\la \ii,\h\ra\neq 0\\\la \ii,\uu_\gl\ra=0}}n_\ii=\sum_{\substack{\la \ii,\bbb\ra\neq 0\\\la \ii,\uu_\gl\ra=0}}n_\ii\leq \sum_{\substack{\la \ii,\bbb\ra= 0\\\la \ii,\uu_\gl\ra\neq0}}n_\ii=\sum_{\substack{\la \ii,\bbb\ra= 0\\\la \ii,\h\ra\neq0}}n_\ii
\eeql
So by combining (\ref{LEtnuheq}) and (\ref{LEtnuhint}) and recalling that $\nu(\h)=1+\nu(\h')$, we conclude that
\beql{LEnuheqfin}
t_{1+\nu(\h')}=t_{\nu(\h)}\leq \sum_{\substack{\la \ii,\h\ra\neq 0\\\la \ii,\bbb\ra=0}}n_\ii+\sum_{\gl\in \bbF_q^*} \sum_{\substack{\la \ii,\bbb\ra= 0\\\la \ii,\h\ra\neq0}}n_\ii=q\sum_{\substack{\la \ii,\bbb\ra= 0\\\la \ii,\h\ra\neq0}}n_\ii,
\eeql
hence
\beql{LEindineq}\sum_{\substack{\la\ii,\bbb\ra= 0\\\la\ii,\h\ra\neq 0}}n_\ii
\geq \lceil t_{1+\nu(\h')}/q\rceil
\eeql
holds for all $\h'\in \cP_{k-1}$. 

Now we set up a 1-1 correspondence between vectors $\ii'\in \bbF_q^{k-1}$  and vectors $\ii\in \bbF_q^k$ for which $\la \ii,\bbb\ra=0$. Writing $\bbb=(1,\bbb')$, we let $\ii'$ correspond with $\ii=(i_1,\ii')$, with $i_1:=-\la \ii',\bbb'\ra$. Note that then 
\beql{LEip}\la \ii,\h\ra=\la \ii',\h'\ra.
\eeql
Furthermore, if $\ii\in \Fqkstar$ with $\la\ii,\bbb\ra=0$ 
corresponds with $\ii'$ in~$\Fqkonestar$,
then we define $n'_{\ii'}:=n_\ii$. Finally, we set $T':=(t_1', \ldots, t'_{k-1})$, where $t'_j:=\lceil t_{j+1}/q\rceil$. Then from (\ref{LEindineq}) and (\ref{LEip}), we find that
\beql{LEfinind}
\sum_{\la\ii',\h'\ra\neq 0}n'_{\ii'}
\geq t'_{\nu(\h')}
\eeql
holds for all $\h'\in \cP_{k-1}$. 
Hence by induction, we have that
\beql{LEG2} \sum_{\la\ii,\bbb\ra=0} n_\ii=\sum_{\ii'\in\Fqkonestar}n'_{\ii'}= n'
\geq \sum_{j=1}^{k-1}\lceil \lceil t_{j+1}/q\rceil/q^{j-1} \rceil=\sum_{j=2}^{k} \lceil t_{j}/q^{j-1} \rceil.
\eeql
By combining (\ref{LEG0}) and (\ref{LEG2}),
we now find that $n\geq \sum_{j=1}^{k}\left\lceil \frac{t_j}{q^{j-1}}\right\rceil$. So the claim for dimension $k$ follows.
\epf
\section{\label{LSuni}The Griesmer bound for linear codes and for linear UEP-codes from the ILP problem}
The Griesmer bound for linear codes can also be proved by our ILP argument. To see that, assume that $\GG=[\g_1, \ldots, \g_n]$ is a $k \times n$ matrix over $\bbF_q$ that generates a $k$-dimensional $q$-ary linear code of length~$n$ with minimum distance $d$. Suppose that 
$\GG$ has $n_\ii$ columns equal to $\ii$ ($\ii\in\bbF_q^k$). Let~$\h^\perp$ be a hyperplane, where $\h\in \bbF_q^k\setminus\{\bfnull\}$. Consider $\cc^\top=\h^\top \GG$. Then $c_j=0$ if and only if $\h^\top \g_j=0$, so $w(\cc)=\sum_{\la\h,\ii\ra\neq 0} n_\ii$.
We conclude that
\[ \sum_{\la\h,\ii\ra\neq 0} n_\ii \geq d\]
holds for every $\h\in \bbF_q^k\setminus \{\bfnull\}$.
So a linear code with generator matrix $\GG$ has minimum distance at least~$d$ if and only if every hyperplane contains at most $n-d$ columns of~$\GG$, or equivalently, if there are at least $d$ columns of~$\GG$ outside every hyperplane.
This establishes the ILP (\ref{LEILP}) for the case where $t_i=d$ for all~$i\in [k]$, hence shows that the Griesmer bound holds for linear codes. 

The Griesmer bound for UEP codes (see \cite[page 23]{Gil88}) can also be obtained from the ILP (\ref{LEILP}). Indeed, suppose that the linear UEP code is generated by a $k\times n$ matrix $\GG$  over~$\bbF_q$. 
Then the separation vector $(s_1, \ldots, s_k)$ of the code is given by 
\[s_j=s_j(\GG)=\min\{w(\h^\top \GG) \mid h_j\neq 0\},\]
($j\in [k]$) \cite{Gil88}. Suppose that the rows of~$\GG$ are ordered in such a way that $s_1\geq  \cdots \geq s_{k}$. Suppose furthermore that $\GG$ has $n_\ii$ columns equal to $\ii$ ($\ii\in\bbF_q^k$). Then if $\h\in \bbF_q^k$ and $h_j=1$ and $h_1= \cdots=h_{j-1}=0$ (so if $\h\in \cP_k$ and $\nu(\h)=j$), then 
\[
\sum_{\la\ii,\h\ra\neq 0}n_\ii = |\{l \mid \h^\top \g_l\neq 0\}| 
=w(\h^\top\GG)\geq s_j= s_{\nu(\h)},
\]
so again we obtain the ILP (\ref{LEILP}); hence the Griesmer bound for UEP-codes also follows from the ILP bound. Conversely, it is not difficult to see that if $(n_\ii)_{\ii\in \bbF_q\setminus\{\bf 0\}}$ is a feasible solution to the ILP (\ref{LEILP}), then with $n=\sum n_\ii$, the $k\times n$ matrix $G$ that has $n_\ii$ columns equal to $\ii$ for all $\ii\in\bbF_q^n\setminus\{\bf 0\}$ satisfies $s_j(\GG)\geq t_j$ for all~$j$. So the problem of finding a linear UEP code with the smallest length $n$ for which $\s\geq (t_1, \ldots, t_k)$ is in fact equivalent to the ILP problem (\ref{LEILP}).
%
%
%
\section{\label{LSopen}Open problems}
1. It would be interesting to find a constructive proof of the Griesmer bound for UDD PIR codes, along the lines of the usual proofs for Griesmer-type bounds for linear codes. So the question is,  given a linear $q$-ary $(t_1, \ldots, t_k)$-PIR code of length~$n$ with $t_1\geq \cdots\geq t_k$, can we construct a $( \lceil t_2/q\rceil, \ldots, \lceil t_k/q\rceil)$-PIR code of length~$n-t_1$?

2. Earlier we mentioned that linear UEP codes have an optimal generator matrix. This matrix can easily be constructed and thus the separation vector of the code can be determined relatively easily. Do PIR codes, and, more generally,  UDD PIR codes, also have an optimal generator matrix? Does the optimal generator matrix for the code, considered as a UEP code, always provide the optimal encoder for the code as an UDD PIR code?
\section{\label{LScon}Conclusions}
Unequal Error Protection (UEP) error-correcting codes were designed for the scenario where some parts of the encoded data need more protection than other parts. The correction properties of an encoder for an UEP code are captured by a generalization of the minimum distance called the {\em separation vector\/}. 
In this paper, we investigate Unequal Data Demand (UDD) PIR codes, generalizing the notion of a $t$-PIR code to include scenarios where some parts of the encoded data are in higher demand than other parts. First we have proved a generalized distance bound for UDD PIR encoders in terms of the separation vector of the associated UEP code. This bound has been used to derive a Griesmer-type bound for linear UDD PIR codes from the corresponding Griesmer bound for linear UEP codes. For an alternative proof of this Griesmer-type bound, we have derived an Integer Linear Programming (ILP) bound for the minimum length of a linear UDD PIR code, and we have determined a lower bound for the optimal solution of this ILP. In addition, we show that this ILP bound can be used to give a uniform proof for the Griesmer bound for linear codes, for linear UEP codes, and for linear UDD PIR codes.


\subsubsection*{Acknowledgments} 
This research was supported by the Estonian Research Council grants PRG49 and PSG114, and by the European Regional Development Fund via CoE project EXCITE.


%
%
%

\begin{thebibliography}{9}
%
\bibitem{CKG+95}
B.~Chor, E.~Kushilevitz and O.~Goldreich, M.~Sudan.
\newblock Private information retrieval.
\newblock in: {P}roc.\ 36-th\ {IEEE Symp.\ on Foundations of Computer Science (FOCS)}, 41--50, 1995
%
\bibitem{DR78}
L.A.~Dunning, W.~Robbins.
\newblock Optimal encodings of linear block codes for unequal error protection.
\newblock {\em Inform.\ Control}, 37:150--177, 1978


\bibitem{FVY15}
A.~Fazeli, A.~Vardy, E.~Yaakobi.
\newblock Codes for distributed {PIR} with low storage overhead.
\newblock In: {Proc.\ IEEE Symp.\ Information Theory ({ISIT})},
  2852--2856, Hong Kong (2015)


\bibitem{Gil88}
W.J.~van Gils.
\newblock Design of error-control coding schemes for three problems of
  noisy information transmission, storage and processing.
\newblock PhD thesis, Eindhoven University of Technology, 1988.
\newblock \url{https://doi.org/10.6100/IR274904}


\bibitem{HL22}H.D.L.~Hollmann, U.~Luha\"a\"ar.
\newblock Optimal possibly nonlinear 3-PIR codes of small size.
\newblock in: Arithmetic of Finite Fields, Proceedings WAIFI 2022, LNCS 13638, Chapter 9

\bibitem{Lin99}
J.H.~van~Lint.
\newblock Introduction to Coding Theory (3ed). 
\newblock Springer, 1999
%

\bibitem{LR18}
H.-Y.~Lin, E.~Rosnes.
\newblock Lengthening and extending binary private information retrieval codes.
\newblock In: {Proc.\ International Zurich Seminar on Information and
  Communication (IZS)}, 113 --117, ETH Zurich. February 21--23 (2018)

\bibitem{MS77}
F.J.~MacWilliams, N.J.A.~Sloane.
\newblock The Theory of Error-Correcting Codes.
\newblock North Holland, 1977
%

\bibitem{LS14}
H.~Lipmaa, V.~Skachek.
\newblock Linear batch codes.
\newblock In: {Proc.\ 4th International Castle Meeting on Coding Theory and
  Applications (ICMCTA)}, 245--253, Palmela, Portugal (2014)


\bibitem{KY21}
S.~Kurz, E.~Yaakobi.
\newblock {PIR} codes with short block length.
\newblock {\em Des.\ Codes, Cryptogr.}, 89:559--587, June 2021

\bibitem{Pus22}M.~Pu\v{s}kin, 
\newblock On Unequal Data Demand Private Information Retrieval Codes.
\newblock Bachelor Thesis, University of Tartu, 2022

\bibitem{Ska18}
V.~Skachek.
\newblock Batch and {PIR} codes and their connections to locally repairable codes.
\newblock In:  Greferath, M., Pavčević, M.~O., Silberstein, N., \'Angeles
  Vázquez-Castro, M.\ (eds) {Network Coding and Subspace Designs}, 
  427--442. Springer (2018)

\bibitem{var-you}
A.~Vardy.
\newblock {Private Information Retrieval: Coding instead of Replication}.
\newblock Talk at the Institate Henri Poincar\'e, March 25 (2016).
\newblock Available at
  \url{https://www.youtube.com/watch?v=WU2-6Da8IyE&t=934s}

\bibitem{ZS15}
J.~Zumbr{\"a}gel, V.~Skachek.
\newblock Talk: On bounds for batch codes.
\newblock Algebraic Combinatorics and Applications (ALCOMA), March 15--20 (2015)







\end{thebibliography}
%

\end{document}